\documentclass{elsart1p}

\usepackage{graphicx}
\usepackage{amssymb}

\def\be{\begin{equation}}
\def\ee{\end{equation}}
\def\bea{\begin{eqnarray}}
\def\eea{\end{eqnarray}}
\def\bear{\begin{array}}
\def\ear{\end{array}}
\def\bfig{\begin{figure}}
\def\efig{\end{figure}}
\def\bcen{\begin{center}}
\def\ecen{\end{center}}
\def\bi{\begin{itemize}}
\def\ei{\end{itemize}}

\def\bra#1{\left\langle #1\right|}
\def\ket#1{\left| #1\right\rangle}

\def\Ket#1{\bigl| #1\bigr\rangle}

\def\chic{\scriptscriptstyle}

\begin{document}

\begin{frontmatter}


\title{Study of the core-excited Fano Resonances in the neutron-rich C-isotopes}
\author[INFN,UG]{S.E.A. Orrigo},
\ead{orrigo@lns.infn.it}
\author[UG]{H. Lenske},
\author[INFN]{F. Cappuzzello},
\author[INFN]{A. Cunsolo}
\address[INFN]{INFN Laboratori Nazionali del Sud and Universit\'a~di Catania, Via S.Sofia 62-64, Catania, Italy}
\address[UG]{Institut f\"ur Theoretische Physik, Universit\"at Giessen, Heinrich Buff-Ring 16, Giessen, Germany } 

\begin{abstract}
Fano Resonances are a new continuum excitation mode in light exotic nuclei. By the QPC model we explain these sharp and asymmetric structures as due to the coupling of a single particle elastic channel to the closed core-excited channels. Theoretical model calculations are compared for $^{15}$C, $^{17}$C and $^{19}$C. The $^{15}$C results are in qualitative agreement with the experimental spectrum.
\end{abstract}

\end{frontmatter}

Mean-field approaches are no longer appropriate in the new scenario of light exotic nuclei. The proximity of the continuum and the increased effects of the correlations are responsible of new excitation modes. The coupling of the 1-QP (one-quasiparticle) and 3-QP core-excited components leads to long-living resonances in the low-energy continuum~\cite{Orrigo:2006rd}. These Bound States Embedded in the Continuum (BSEC), observed by us in $^{11}$Be~\cite{CapPLB01,CapNPA04} and $^{15}$C~\cite{OrrigoVAR03,OrrigoPhD04,OrrigoEPL04}, can be regarded into the more general class of the Fano Resonances~\cite{FanoPR61} and are expected to be important in light neutron-rich nuclei. The systematic comparison for nuclei with slightly different neutron separation energies $S_{n}$ provides information on the evolution of the phenomenon when going towards more extreme conditions. A strong enhancement of the BSEC excitations is predicted in the C-isotopes.

Our approach~\cite{Orrigo:2006rd} to the Fano resonances in exotic nuclei extends the quasiparticle-core coupling (QPC) model~\cite{LenskePPNP04} in the continuum and incorporates information from HFB and QRPA calculations. The QPC Hamiltonian includes the operators acting on the 1-QP and 3-QP components, i.e., single particle (s.p.) motion with inert core and 2-QP QRPA core excitations. The residual interaction $V_{13}$ couples the 1-QP and 3-QP states. The QPC eigenstates are a superposition of 1-QP $\ket{\,n_J}$ and core-excited 3-QP $\ket{\,\left( j'J_C\right)_J }$ states:
\be
\Ket{\varphi_J} = \sum_n z_n(E)\Ket{\,n_J} + \int d\epsilon \, z_\epsilon(E)\Ket{\,\epsilon_J} 
+ \sum_{j'\,J_C} z_{\chic j'J_C}(E)\Ket{\,\left( j'J_C\right)_J }  
\ee

The first term fixes the total spin $J$ and parity, the second term accounts for the s.p. continuum (energy $\epsilon_1>0$), the third term for the core excitations. The BSEC include bound core-excited states ($\epsilon_i<0$), hence for BSEC-coupling the flux of the elastic component is asymptotically conserved, fully accounting for channel coupling and mixtures between the discrete and continuum parts of the spectrum~\cite{Orrigo:2006rd}. Projecting the Schr\"odinger equation onto the 1-QP and 3-QP components and integrating we obtain $N$ coupled equations which are solved numerically. The radial wave functions are given by superposition of the eigenfunctions $\chi_{im}(r)=a_{im}(r)j_{l_i}(Q_mr)$ for radii $r < R_A$ (internal region) and by their asymptotic forms for $r \geq R_M \gg R_A$ ($R_M$ = matching radius):
\be
u_{i,l_i}(r)=\sum^N_{m=1} b_{m} \,\chi_{im}(r); \quad
u_{i,l_i}(r)=j_{l_1}(k_1r) \,\delta_{li} + C_{li} \,h_{l_i}^{(+)}(k_ir); \quad i=1,\ldots,N
\ee

The coefficients $b_m$ and $C_{1i}$ are determined by matching the internal and asymptotic solutions, with incoming and outgoing waves present only in the g.s. elastic channel since all other channels are asymptotically closed. The scattering matrix elements in the open channel determine the partial wave elastic cross sections $\sigma_{ii} = \sigma_i(k_i)$: 
\be
\sigma_i(k_i)=\frac{4 \pi}{k_i^2} \frac{2j+1}{2s+1} \left| C_{ii}(k) \right|^2
\ee

The model is applied to $^{15}$C, $^{17}$C, $^{19}$C ($S_{n} =$ 1.218, 0.73, 0.16~MeV, respectively) and used to explain the structures above the particle threshold observed in $^{15}$C~\cite{OrrigoVAR03,OrrigoPhD04,OrrigoEPL04}. HFB potentials are used as reference to reproduce the single neutron states in the core systems $^{14}$C, $^{16}$C, $^{18}$C. The particle-core interactions are described by the coupling form factors $F_{J_C}^{(i)}=\bra{0}V_{13}\ket{J_C}=\beta_{J_C}^{(i)}F_0$, where $F_0$ is a free parameter and $\beta_{J_C}^{(i)}$ are state dependent core transition amplitudes, taken from the $^{14}C$($\alpha$,$\alpha$') reaction~\cite{AjNPA91} for the $^{14}$C core and calculated by QRPA for $^{16}$C and $^{18}$C. The n+C-core partial wave elastic cross sections $\sigma_{11}$ are calculated. When the interaction is turned on $(F_0 \neq 0)$ narrow and asymmetric resonances appear in the d$_{5/2}$-wave cross section, displayed in Fig.1 for $^{15}$C (a), $^{17}$C (c) and $^{19}$C (d). The s- and p-wave cross sections give rise to a continuous background~\cite{Orrigo:2006rd}.
\bfig[th!]
\label{Fig1}
\begin{minipage}{.49\linewidth}
\bcen
\includegraphics[width=\textwidth]{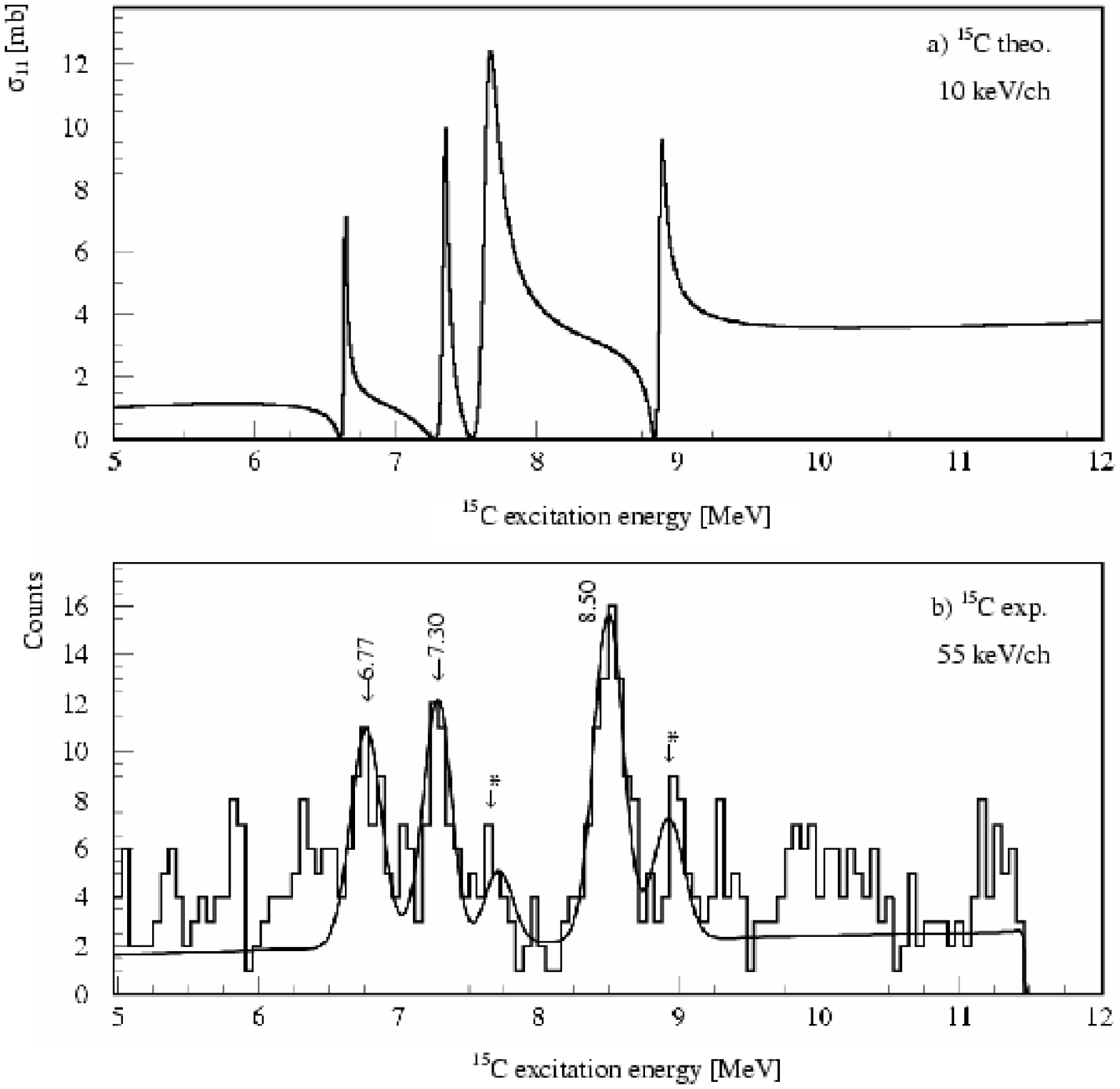}
\ecen
\end{minipage}
\hfill
\begin{minipage}{.49\linewidth}
\bcen
\includegraphics[width=0.72\textwidth]{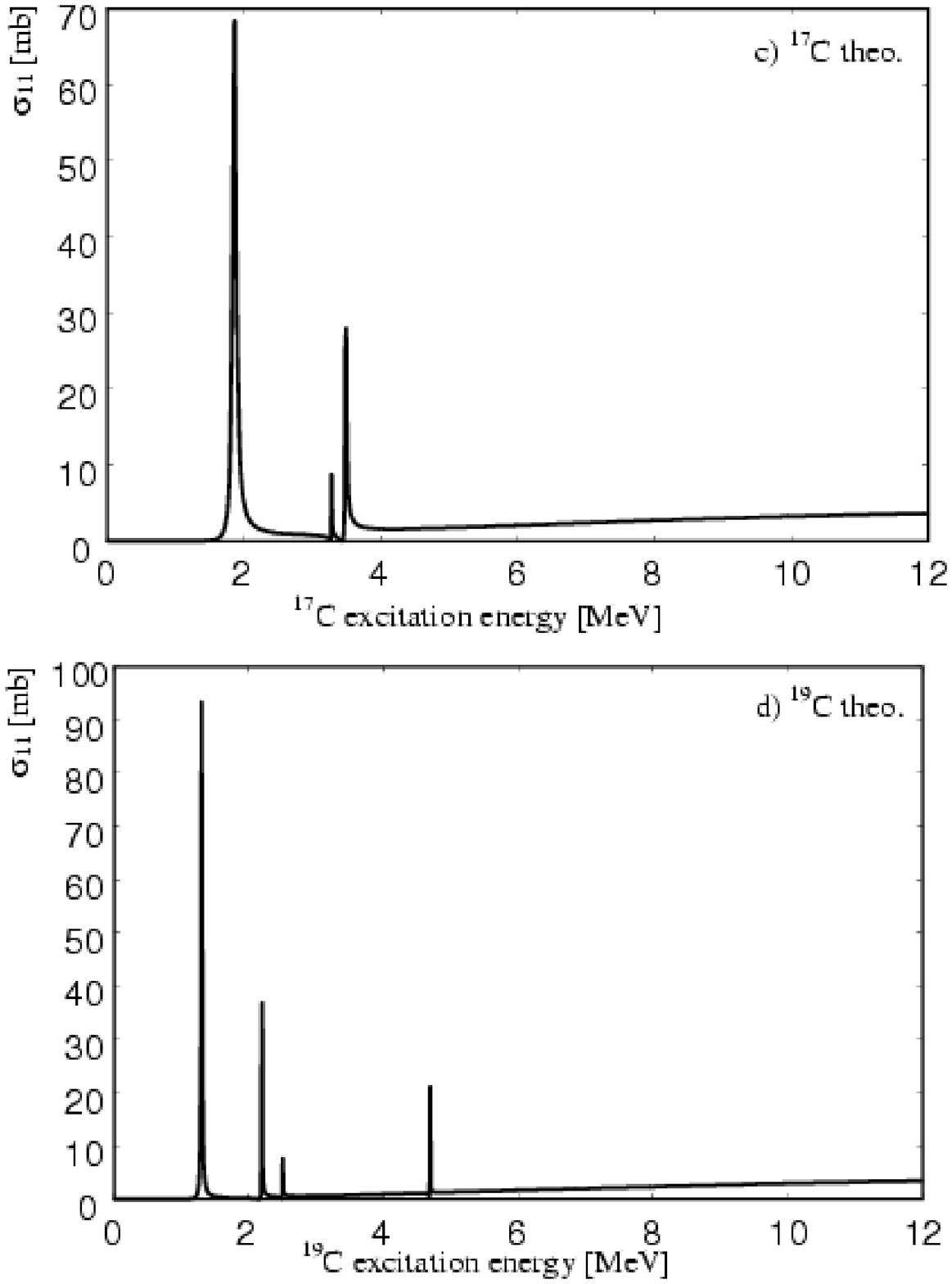}
\ecen
\end{minipage}
\caption{d$_{5/2}$-wave elastic cross section $\sigma_{11}$ for $^{15}$C (a), $^{17}$C (c), $^{19}$C (d). In (b) the excitation energy spectrum for the $^{15}$N($^{7}$Li,$^{7}$Be)$^{15}$C reaction at $E_{inc} = 55$~MeV and $\theta_{lab} = 14^\circ$ is shown~\cite{OrrigoVAR03,OrrigoPhD04,OrrigoEPL04}.} 
\efig

The calculations for $^{15}$C include the continuum and core-excited states $E_C (J^\pi) =$ 6.094 ($1^-$), 6.728 ($3^-$), 7.012 ($2^+$), 8.317 ($2^+$)~MeV. Several long-living resonance structures above the particle threshold appear in the region 6-10~MeV. Their asymmetric line shape results from bound-continuum coupling, typical for Fano resonances~\cite{FanoPR61}. The theoretical $\sigma_{11}$ (Fig.1a) is compared on a qualitative level to the $^{15}$C experimental spectrum (Fig.1b). The calculations do not account for reaction dynamics, however describe the rescattering of the neutron after production on the residual nucleus, responsible for the structures seen in the measured spectrum. Four theoretical levels are found at $E_x =$ 6.67, 7.36, 7.70, 8.92~MeV of widths $\Gamma =$ 66, 80, 141, 85~keV. The experimental resonances are observed at $E_x =$ (6.77 $\pm$ 0.06, 7.30 $\pm$ 0.06, 8.50 $\pm$ 0.06)~MeV of widths $\Gamma =$ 160, 70, 140~keV. The theoretical and experimental values are very close. The asymmetry of the 8.5~MeV peak, signature of BSEC excitation, is an experimental evidence of core polarization effects.

For $^{17}$C (Fig.1c) we include the core-excited states $E_C (J^\pi) =$ 1.766 ($2^+$), 3.986 ($2^+$), 4.142 ($4^+$)~MeV. For $^{19}$C (Fig.1d), the states $E_C (J^\pi) =$ 1.620 ($2^+$), 2.967 ($4^+$), 3.313 ($2^+$), 5.502 ($1^-$)~MeV, all from QRPA calculation except for 1.620~MeV. A progressive lowering of the BSEC energies and an enhancement in the intensity are seen when going from $^{15}$C to $^{17}$C and $^{19}$C. These are clear indications of an enhanced effect of dynamical correlations with increasing neutron excess, which shift more and more strength from the continuum to the BSEC excitations. So for more exotic systems sharper and more intense BSEC are expected at lower $E_x$, making the experimental observation more accessible.

An important prediction of Fano's theory~\cite{FanoPR61} is the direct relation between the coupling matrix elements and the width of a Fano resonance. This is also found here when varying the interaction strength to our will: for $F_0\sim$1-2~MeV the coupling is weak and gives only small fluctuations on the continuous background; for intermediate $F_0$ the resonances become more intense and narrow; from $F_0\sim7$~MeV the width increases until individual states are no longer resolved. The results shown, with well resolved states and visible interferences, are for $F_0 = 5$~MeV, a rather good compromise. Hence, measurements of resonance line shapes will provide information on residual interactions in exotic nuclei.

Since the BSEC phenomena lead to a change of the level density close to the continuum threshold, such structures will be highly important in the astrophysical capture and knockout reactions contributing to nuclear synthesis in neutron-rich stellar environments.



\end{document}